\documentstyle[12pt]{article}

\begin{document}
\setlength{\baselineskip}{15pt}

\title{Hyperfinite-Dimensional Representations of Canonical Commutation
Relation}
\author{Hideyasu Yamashita\\{\it Graduate School of Human Informatics,} \\
{\it Nagoya University, Nagoya 464-01, Japan}}
\date{}

\maketitle
\sloppy

\begin{abstract}
This paper presents some methods of representing canonical commutation
relations in terms of hyperfinite-dimensional matrices, which are constructed
by nonstandard analysis. The first method uses representations of a nonstandard
extension of finite Heisenberg group, called hyperfinite Heisenberg
group. The second is based on hyperfinite-dimensional representations
of $so(3)$. Then, the cases of infinite degree of freedom are argued in
terms of the algebra of hyperfinite parafermi oscillators, which is 
mathematically equivalent to a hyperfinite-dimensional representation
of $so(n)$.\vspace{4mm}
\end{abstract}

\def\N{{\bf N}}
\def\R{{\bf R}}
\def\C{{\bf C}}
\def\Z{{\bf Z}}
\def\M{{\bf M}}
\def\P{{\cal P}}
\def\us{\!{}^{\star}}
\def\H{{\cal H}}
\def\K{{\cal K}}
\def\fin{\mathop{\rm fin}\nolimits}
\def\dag{{}^{\dagger}}
\def\dom{\mathop{\rm Dom}\nolimits}
\def\hull{\widehat{\!{}^{\star}{\bf C}^\nu}}
\def\st{\mathop{\rm st}\nolimits}
\def\tr{\mathop{\rm Tr}\nolimits}
\def\F{{\cal F}}

\newpage
\section*{I. INTRODUCTION}

In the field of mathematical foundations of quantum physics, especially
of infinite degree of freedom, it is often argued that we need any 
extensions of separable Hilbert space as the fundamental mathematical 
tools, such as the linear space of generalized functions.

In recent years, another alternative, {\it nonstandard-analytical
extension} method, is developed by some authors${}^{1,2,3,4}$. The present 
paper adopts an approach in use of 
hyperfinite-dimensional linear spaces, constructed by nonstandard 
analysis${}^{5,6}$ (see Appendix to find the basic definitions).
 
This approach also is closely related with that of {\it finite-dimensional
quantum physics}${}^{7,8,9,10}$;
nonstandard extensions of finite-dimensional models in quantum physics
turn out to be the powerful tools for the approach. 

The purpose of the present paper is representing canonical commutation
relasion  (CCR) $[Q,P]=i$ (let us take the natural unit $\hbar=1$), or 
equivalently,
$[a,a^\dagger]=1$ where $a=(Q+iP)/\sqrt 2$, in a hyperfinite-dimensional
linear space. Let $\nu$ be an infinite hypernatural number, and $Q$ and $P$  
internal $\nu\times\nu$ hermitian matrices. $Q$ and $P$ act on the 
$\nu$-dimensional internal linear space $\us\C^\nu$ as internal linear 
operators. No standard finite-dimensional matrices satisfy CCR. 
By the transfer principle in nonstandard analysis,
the fact also holds for any internal hyperfinite-dimensional matrices,
so we have $[Q,P]\neq i$. Thus we must arrange the standard CCR for 
hyperfinite-dimensional matrices. First we consider the following weaker
version of CCR: $[Q,P]\approx i$, that is, $\|[Q,P]-i\|_\infty\approx 0$,
where $\|A\|_\infty$ is given by 
\[\|A\|_\infty=\sup\{\|A\xi\|\,:\,\,\|\xi\|\leq 1, \xi\in\us\C^\nu\}\]
This is equivalent to the condition that for any $\xi\in\us\C^\nu$ with
finite norm, $[Q,P]\xi\approx i\xi$ ( i.e., $\| ([Q,P]-i)\xi\|\approx 0$
where $\|\cdot\|$ is the norm on $\us\C^\nu$ defined by $\|(\xi_1,...,\xi_\nu)
\|=\sqrt{\xi_1^2+\cdots+\xi_\nu ^2}$ ). We see that the condition never holds
as follows. Suppose the condition holds. Then any eigenvalue $\lambda$ of
$[Q,P]$ satisfies $\lambda\approx i$. This contradicts $\tr([Q,P])=0$. 
Now, let us define the 
still weaker notion of representation of CCR by limiting the domain of
the operators $Q$ and $P$.

{\it Definition 1.1:} Let $Q_k,P_k$ be $\nu\times\nu$ internal hermitian
matrices $(k=1,2,...<\infty)$, and $S$ be an
external subspace of $\us\C^\nu$ which consists only of the vectors with
finite norm. The set of the triples $(Q_k, P_k,S)$ (or of pairs $(a_k,S)$
where $a_k=(Q_k+iP_k)/\sqrt 2$) is called a {\it quasi-representation of 
CCR\/} if
\[[Q_k,P_l]\xi\approx i\delta_{kl}\xi,\,\,\,
[Q_k,Q_l]\xi \approx [P_k,P_l]\xi\approx 0\]
for any $\xi\in S$ and $k,l=1,2,...<\infty$. A quasi-representation of CCR
 $(Q_k, P_k,S)$ is called a {\it representation
of CCR\/} if $S$ is invariant respect to $Q_k,P_k$ (i.e., $Q_kS,P_kS
\subseteq S$).

Hereafter, we will use the following notations.\\
(1) $\us$ is the mapping of nonstandard extension.\\
(2) $\N (\us\N)$ is the set of natural (hypernatural) numbers.\\
(3) $\Z (\us\Z)$ is the set of integers (hyperintegers).\\
(4) $\R (\us\R)$ is the set of real (hyperreal) numbers.\\
(5) $\C (\us\C)$ is the set of complex (hypercomplex) numbers.

\section*{II. CANONICAL PAIR OF UNITARY MATRICES}
\def\k{{\bf K}}

{\it Definition 2.1:} Let $M(\us\C, \nu)$
denote the set of internal $\nu\times\nu$ matrices ($\nu\in\us\N$). Unitary 
matrices $U,V\in M(\us\C, \nu)$ are called a {\it canonical pair\/} if 
$U^k,V^k\neq 1 (k=1,...,\nu-1)$, $U^\nu=V^\nu=1$ and $UV=e^{2\pi i/\nu}VU$.

Let $|u_0\rangle$ be a normalized eigenvector of $U$ with eigenvalue $c_0$.

{\it Lemma 2.1:} For any $n\in\us\N$, $V^n|u_0\rangle$ is an
eigenvector of $U$, and its eigenvalue is $e^{2\pi ni/\nu}c_0$.

{\it Proof:} Induction for $n$. $UV^{n+1}|u_0\rangle=
UVV^{n}|u_0\rangle=e^{2\pi i/\nu}VUV^n|u_0\rangle=
e^{2\pi i/\nu}Ve^{2\pi ni/\nu}c_0V^n|u_0\rangle=c_0e^{2\pi (n+1)i/\nu}V^{n+1}
|u_0\rangle.$ Q.E.D. 

From the definition, we find that all the eigenvalues
of $U$ are $e^{2\pi i/\nu},e^{4\pi i/\nu},...,e^{2\nu\pi i/\nu}=1$.
Let us assume $c_0=1$.

{\it Lemma 2.2:} Define $|u_k\rangle$ and $|v_n\rangle$ by $|u_k\rangle=
V^k|u_0\rangle$ and $|v_n\rangle =\frac1{\sqrt \nu}\sum_{k=0}
^{\nu-1}e^{2\pi nki/\nu}|u_k\rangle$ $(k,n\in\us\N)$. $|v_n\rangle$ 
is an eigenvector of $V$, and its eigenvalue is $e^{-2\pi ni/\nu}$.

{\it Proof:} $V|v_n\rangle =\frac1{\sqrt \nu}\sum_{k=0}
^{\nu-1}e^{2\pi nki/\nu}V^{k+1}|u_0\rangle =\\
e^{-2\pi ni/\nu}\frac1{\sqrt \nu}
\sum_{k=0}^{\nu-1}e^{2\pi nki/\nu}V^{k}|u_0\rangle =e^{-2\pi ni/\nu}
|v_n\rangle.$ Q.E.D.

From these lemmas, we see that $|u_k\rangle$ and $|v_k\rangle \,\,
(k=0,...,\nu-1)$ are complete orthonormal systems of $\us\C^\nu$. Thus we have 
\[V=\sum_{k=0}^{\nu-1}|u_{k+1}\rangle\langle u_k|=
\sum_{k=0}^{\nu-1}e^{-2\pi ni/\nu}|v_n\rangle\langle v_n|\]
\[U=\sum_{k=0}^{\nu-1}|v_{k+1}\rangle\langle v_k|=
\sum_{k=0}^{\nu-1}e^{2\pi ni/\nu}|u_n\rangle\langle u_n|\]

The following proposition follows from these consequences.

{\it Proposition 2.3:} The canonical pair of $\nu\times\nu$ unitary matrices
is unique in the sense of unitary equivalence, that is, if $(U,V)$ and
$(U',V')$ are two such canonical pairs, then there is a unitary matrix $W$
such that $U'=WUW^\dagger, V'=WVW^\dagger$.

{\it Definition 2.2:} Let $A\in M(\us\C,\nu)$. $|a\rangle \in
\us\C^\nu$ is called {\it approximately invariant\/} vector of $A$ iff
$\langle a|a\rangle<\infty$ and $A|a\rangle\approx |a\rangle$.

Let $I(A)$ denote the set of all the approximately invariant vectors of $A$.
$I(A)$ is an external subspace of $\us\C^\nu$.
 
{\bf Theorem 2.4:} If $\nu >\infty$, then $I(U)\cap I(V)$ is an invariant 
subspace of $U$ and $V$, and is infinite-dimensional (i.e., it has an
infinite orthonormal system), and 
\[\frac\nu{2\pi mn}[U^m,V^n]|a\rangle\approx i|a\rangle.\]
holds for any $|a\rangle\in I(U)\cap I(V)$ and $m,n\in\Z$.

{\it Proof:} $I(U)\cap I(V)$ is clearly an invariant subspace.
Suppose $l=0,1,...<\infty,\,\,\,\, \mu>\infty$ 
and $\mu/\nu\approx 0$. Define $|l\rangle$ by 
\[|l\rangle =\frac1{\sqrt{\mu}}\sum_{k=l\mu}^{(l+1)\mu-1}|u_k\rangle\]
$\{|0\rangle, |1\rangle,...\}$ is an orthonormal system of $I(U)\cap I(V)$.
In fact, $\|V|l\rangle -|l\rangle \|^2=(1/\mu)\||u_{(l+1)\mu}\rangle -
|u_{l\mu}\rangle\|^2\approx 0$, and $ \|U|l\rangle-|l\rangle \|^2=
(1/\mu)\sum_{k=l\mu}^{(l+1)\mu-1}|e^{2\pi \mu i/\nu}-1|^2\approx 0$.
We have also $(\nu/2\pi mn)(U^mV^n-V^nU^m)|a\rangle=
\nu/{2\pi mn}(e^{2\pi mni/\nu}-1)V^nU^m|a\rangle\approx iV^nU^m|a\rangle\approx
i|a\rangle.$ Q.E.D. 

{\it Corollary:} Suppose $m,n\in\N$. Let $P^{(m)},Q^{(n)}$ be 
hermitian matrices defined by
\[P^{(m)}=\frac i{m}\sqrt{\frac\nu{8\pi}}(U^m-U^{-m}),\,\,\,
Q^{(n)}=\frac i{n}\sqrt{\frac\nu{8\pi}}(V^n-V^{-n})\]
If $|a\rangle\in I(U)\cap I(V)$, then
\[[Q^{(n)},P^{(m)}]|a\rangle\approx i|a\rangle,\]
and hence, $(Q^{(n)},P^{(m)},I(U)\cap I(V))$ is a quasi-representation of CCR
for each $m$ and $n$.
 
We will give some other properties of a canonical pair of unitary matrices.

{\it Definition 2.3:} Let $K\in\N$. Let $\k$ be the ring of residue classes
of $\Z$ modulo $K$, i.e., $\k=\Z/K\Z$. For $k,k'\in \k$, $k\oplus 
k'$ stands for the sum of $k$ and $k'$ in $\k$, $k\otimes k'$ the product of 
$k$ and $k'$, and $\ominus k$ the minus $k$. The {\it finite Heisenberg
group\/} based on $\k$ is the group $H_K$ with the underlying set 
$\k\times\k\times\k$ whose group operation is 
\[(k,l,m)(k',l',m')=(k\oplus k',l\oplus l', m\oplus m'\oplus(k\otimes l')).\]
The notion of {\it Hyperfinite Heisenberg group\/} is the nonstandard 
(internal) extension of that of finite Heisenberg group; its definition 
is given by 
substituting $\us\N$ and $\us\Z$ for $\N$ and $\Z$ in the above definition, 
respectively. 

Suppose $\nu>\infty$. The canonical pair $U,V$ generates an internal group
$G$. We see 
\[G=\{e^{2\pi im/\nu}V^lU^k|\,k,l,m=0,1,...,\nu-1\}\]
The following property is easily seen.

{\it Proposition 2.5:} Define the mapping $\pi:H_\nu\rightarrow M(\us\C,\nu)$
by $\pi(k,l,m)=e^{2\pi im/\nu}V^lU^k$. $\pi$ is an internal irreducible
unitary representation of the hyperfinite Heisenberg group $H_\nu$.

Consider $\R^3$ with coordinates $(p,q,t)$. We make $\R^3$ into a locally
compact group with group law
\[(p,q,t)(p',q',t')=(p+p',q+q',t+t'+pq').\]
We call this group the {\it Heisenberg group\/} and denote it by $H$.
The {\it Sch\"odinger representation\/} of $H$ is the storong continuous
unitary representation $\rho$ of $H$ on $L^2(\R)$ defined by
\[(\rho(p,q,t)f)(x)=e^{2\pi i(t+qx)}f(x+p)\]
for all $f\in L^2(\R)$.
This is a realization of Weyl's commutation relation $U(q)V(p)=
e^{iqp}V(p)U(q)$ where $U(q)$ and $V(p)$ are one-parameter unitary groups.
From the representation of $H_\nu$, Ojima and Ozawa${}^1$ constructed a 
representation of $H$ in a hyperfinite-dimensional
Hilbert space which is unitary equivalent to the Sch\"odinger representation 
of $H$. By using this representation, they found a nonstandard-analytical 
proof of noncommutative Parseval's identity
\[\int\!\!\int_{\R^2}\langle\psi_1|\rho(p,q,0)|\phi_1\rangle
\langle\phi_2|\rho(p,q,0)^{-1}|\psi_2\rangle dp\,dq=
\langle\phi_2|\phi_1\rangle\langle\psi_1|\psi_2\rangle,\]
for all $\psi_i,\phi_i\in L^2(\R)\,(i=1,2)$.

\section*{III. $SO(3)$-REPRESENTATION OF CCR}

Standard position and momentum operators $x,p$ with $[x,p]=i$ have 
the $x$-$p$ rotation covariant property, that is, 
\[x\cos\theta +p\sin\theta =e^{i\theta H}xe^{-i\theta H}\]
where $H$ is given by $x^2+p^2$. This property is essential when 
dealing with quantum harmonic oscillators. On the other hand,  
$Q^{(n)}$ and $P^{(m)}$ defined above have no clear covariant property of 
this sort. This fact reveals a defect of these matrices. In this section, 
we argue another hyperfinite-dimensional representation of CCR,
which have the clear rotation covariance.   

Let $J_1,J_2$ and $J_3$ be an irreducible set of $(p+1)\times(p+1)$ hermitian
matrices which satisfies
\[[J_k,J_l]=i\epsilon_{klm}J_m\]
where $\epsilon_{klm}$ is Levi-Civita symbol. $iJ_1,iJ_2$ and $iJ_3$
generate a $p+1$ dimensional irreducible representation of Lie algebra
$so(3)$. It is known that $J_k$ has eigenvalues
\[\frac{p}2,\,\,\,\,\,\frac{p}2-1,...,\,\,\,1-\frac{p}2,\,\,\,\,-\frac{p}2\]
and corresponding eigenvectors are given by
\[|J_3;\frac{p}2\rangle,\,\,\,\,J_-|J_3;\frac{p}2\rangle,...,\,\,\,\,J_-^p
|J_3;\frac{p}2\rangle\]
where $|J_3;\frac{p}2\rangle$ is a normalized eigenvector with eigenvalue
$p/2$ and $J_-=J_1-iJ_2$. Let $j=p/2$ and $|J_3;m\rangle=J_-^{j-m}
|J_3;j\rangle /\|J_-^{j-m}|J_3;j\rangle\|$. It is shown that
\[\langle J_3;m'|J_-|J_3;m\rangle=\delta_{m'\, m-1}\sqrt{(j+m)(j-m+1)}\]
We consider the nonstandard version of this representation;
let $p$ be an infinite hypernatural number, and $J_k$ be the internal
matrices satisfying the above commutation relations. Define $Q$ and $P$ by
$Q=J_1/\sqrt j,\,\,P=J_2/\sqrt j$.
 
{\bf Theorem 3.1:} Let $S\subset\us\C^\nu$ be the subspace 
finitely spanned by $|J_3;j-k\rangle,\,\,\,(k=0,1,...<\infty)$, i.e., 
\[S=\{\sum_{k=0}^nc_k|J_3;j-k\rangle\,:\,\,c_k\in\us\C, |c_k|<\infty, 
n\in\N\}\]
$S$ is an invariant subspace of $P$ and $Q$, and
\[[Q,P]|\xi\rangle\approx i|\xi\rangle\]
for each $|\xi\rangle\in S$, and hence $(Q,P,S)$ is a representation of CCR.

{\it Proof:} The invariance of $S$ is seen from
\[Q|J_3;j-k\rangle=\frac1{\sqrt j} \frac1{2}(J_-+J_-^\dagger)
|J_3;j-k\rangle\]
\[=\frac1{2}\sqrt{(2-k/j)(k+1)}|J_3;j-k-1\rangle+\frac1{2}
\sqrt{k(2-k/j+1/j)}|J_3;j-k+1\rangle\in S\]
\[P|J_3;j-k\rangle=\frac1{\sqrt j} \frac{i}2(J_-^\dagger-J_-)|J_3;j-k\rangle\]
\[=\frac{i}{2}\sqrt{(2-k/j)(k+1)}|J_3;j-k-1\rangle
-\frac i{2}\sqrt{k(2-k/j+1/j)}|J_3;j-k+1\rangle\in S\]
And we have $[Q,P]|J_3;j-k\rangle=\frac{i}{j}J_3|J_3;j-k\rangle=
i(1-k/j)|J_3;j-k\rangle\approx i|J_3;j-k\rangle$. Q.E.D. 

We find the rotation covariance of these operators;
\[Q\cos\theta+P\sin\theta=e^{i\theta J_3}Qe^{-i\theta J_3}.\]
Moreover, it turns out that this representation is convenient in dealing
with coherent states. Let $R(\theta,\phi)$ denote the rotation matrix
\[R(\theta,\phi)=e^{i\theta(J_1\sin\phi-J_2\cos\phi)}\]
$R(\theta,\phi)$ has also the expression${}^{11}$
\[R(\theta,\phi)=e^{\mu J_-}e^{-\log(1+|\mu|^2)J_3}e^{-\mu^*J_-^\dagger}\]
where $\mu=e^{i\phi}\tan\frac{\theta}{2}$.
Define $|\theta, \phi\rangle$ by $|\theta, \phi\rangle=R(\theta,\phi)
|J_3;j\rangle$. It is shown that
\[|\theta, \phi\rangle=\frac1{(1+|\mu|^2)^j}\sum_{k=0}^{2j}{2j\choose k}
^{1/2}\mu^k|J_3;j-k\rangle\]
If $|\mu|\sqrt j<\infty$ and $k<\infty$,
\[\frac1{(1+|\mu|^2)^j}\approx e^{-j|\mu|^2},\,\,\,\,\,\,\,\,
{2j \choose k}^{1/2}\mu^k
\approx\frac{(\sqrt{2j}\mu)^k}
{\sqrt{k!}}\]
Thus we find
\[\langle J_3;j-k|\theta, \phi\rangle\approx e^{-|z|^2/2}\frac{z^k}{k!}\]
where $z=\mu\sqrt{2j}$. Corresponding standard coherent state $|z\rangle$ 
satisfies
\[\langle k|z\rangle=e^{-|z|^2/2}\frac{z^k}{k!}\]
where $|k\rangle=a^{\dagger k}|0\rangle/\|a^{\dagger k}|0\rangle\|$,
 $a^\dagger$ being the standard bose creation operator and $|0\rangle$ 
the vacuum state.

\section*{IV. REPRESENTATIONS OF $SO(n)$}

The hyperfinite-dimensional representation of CCR which has more 
general rotation
covariance is constructed by hyperfinite-dimensional representation of
$so(n)$. This section reviews the method of {\it spin representation\/} in
use of Clifford algebra. Let $C(\R_-^n)$ be the $\R$-vector space 
the base of which is the $2^n$ elements
\[1,\,\,e_1,...,e_n,\,\,\,e_1e_2,...,e_ie_j\,\,(i<j),...,
e_{n-1}e_n,...,\]
\[e_{i_1}\cdots e_{i_k}\,\,\,(i_1<\cdots <i_k),...,e_1\cdots e_n \]
Define the product of these elements by
\[(e_{i_1}\cdots e_{i_k})(e_{j_1}\cdots e_{j_l})=e_{i_1}\cdots e_{i_k}
e_{j_1}\cdots e_{j_l}\]
and require the relation
\[e_i^2=-1,\,\,\,(i=1,...,n),\,\,\,e_ie_j=-e_je_i\,\,(i\neq j),\,\,\,
(\lambda 1)e_i=e_i(\lambda 1)\,\,\,(\lambda\in\R)\]
Assume that $1$ is the unit of  the product, and that the product is 
associative. Now, $C(\R_-^n)$ is $2^n$-dimensional $\R$-algebra.
$C(\R_-^n)$ is called {\it Clifford algebra}. 

Let us identify $(\lambda_1,...,\lambda_n)\in\R^n$ with $\sum_{k=1}^n
\lambda_ke_k$. Then we have $S^{n-1}\subset\R^n\subset C(\R_-^n)$
($S^{n-1}$ is the unit sphere in $\R^n$). The following fact is shown.

{\bf Theorem 4.1:} $Spin(n)$ defined by
\[Spin(n)=\{\alpha\in C(\R_-^n)\,|\,\,\,\alpha=a_1\cdots a_m,\,a_i
\in S^{n-1},\, m=2,4,...\}\]
is a topological group, and isomorphic to the covering group of $SO(n)$.
$e_{ij}=e_ie_j\,\,(i,j=1,...,n)$ is a base (of a representation) of $so(n)$.

If $n=2\nu+1$, $e_1,...,e_n$ is represented by $2^\nu\times 2^\nu$ matrices
as follows. Let $V_k\simeq\C^2\,\,(k=1,...,\nu)$. 
 Pauli matrices $\sigma_{1,k},\,\sigma_{2,k},\,\sigma_{3,k}$ that act on  
$V_k$ are represented as
\[\sigma_{1,k}=\pmatrix{0&1\cr1&0},\,\,\,\,
\sigma_{2,k}=\pmatrix{0&-i\cr i&0},\,\,\,\,
\sigma_{3,k}=\pmatrix{1&0\cr 0&-1}.\]
Define $V$ by
\[V=V_1\otimes\cdots\otimes V_\nu,\]
 and define $\hat{\sigma}_{c,k}$ and $\gamma_{i}$
that acts on $V$ by
\[\hat{\sigma}_{c,k}=\overbrace{1\otimes\cdots\otimes1}^{k-1}
\otimes{\sigma_{c,k}}\otimes 1\otimes
\cdots\otimes 1,\,\,\,\,\,c=1,2,3,\]
\[\gamma_{2k-1}=\hat{\sigma}_{2,k}\hat{\sigma}_{3,k+1}
\cdots\hat{\sigma}_{3,\nu},\]
\[\gamma_{2k}=-\hat{\sigma}_{1,k}\hat{\sigma}_{3,k+1}
\cdots\hat{\sigma}_{3,\nu},\]
\[\gamma_{2\nu+1}=\hat{\sigma}_{3,1}\hat{\sigma}_{3,2}
\cdots\hat{\sigma}_{3,\nu}.\]

Direct calculations show the following relations.
\[\{\gamma_{i},\gamma_{j}\}=2\delta_{i\,j},\,\,\,\,\,\,
i,j=1,...,2\nu.\]
Thus,we can take $e_j=i\gamma_j$ (this is called the {\it spin representation
\/}).

Moreover we have the tensor product representation of $so(n)$ from one
constructed above; 
\[\tilde e_{ij}=\sum_{l=0}^{p-1}\overbrace{1\otimes\cdots\otimes 1}^{l}
\otimes e_{ij}\otimes \overbrace{1\otimes\cdots\otimes 1}^{p-l-1}\]
also genarate a representation of $so(n)$. 

In the next section, we introduce a hyperfinite representation of $so(n)$,
using these mathematical tools.

\section*{V. HYPERFINITE PARAFERMI REPRESENTATION} 

In this section, we construct a hyperfinite-dimensional representation
of the algebra of standard infinite bose oscillators, that is, 
the algebra generated by bose annihilation operators $a_1,a_2,...$ 
satisfying $[a_j,a_k^\dagger]=\delta_{jk}$ and $[a_j,a_k]=0$,
with the condition that nonzero vector $|0\rangle$ that 
satisfies $a_k|0\rangle=0$ is unique except the scalar multiples
({\it the uniqueness of vacuum\/}).
This representation is constructed by using hyperfinite internal
representations of the algebra of parafermi oscillators, which is
mathematically equivalent to the hyperfinite-dimensional spin 
representation of $so(n)$.
 
{\it Definition 5.1:}${}^{12,13}$ Let $\nu\in\N$, and suppose that for some
$d\in\N$, $b_1,...,b_\nu\in M(d,\C)$ (i.e.,  $b_1,...,b_\nu$ are    
finite-dimensional matrices). $b_1,...,b_\nu$ is called {\it
annihilation operators\/} of parafermi oscillators of order $p\in\N$
if they satisfy
\[[b_k,[b_l\dag,b_m]]=2\delta_{k\,l}b_m\hspace{4em}\]
\[[b_k,[b_l\dag,b_m\dag]]=2\delta_{k\,l}b_m\dag-2\delta_{k\,m}b_l\dag\]
\[[b_k,[b_l,b_m]]=0.\hspace{6em}.\]
and the uniqueness of vacuum $|0\rangle$, and,
\[b_kb_l\dag|0\rangle =\delta_{k\,l}p|0\rangle.\]
$b_1^\dagger,...,b_\nu^\dagger$ are called {\it creation operators\/}
of parafermi oscillators of order $p$.
{\it Hyperfinite annihilation operators\/} of parafermi oscillators are
the internal matrices
 defined in the above definition by substituting $\us\N$ and $\us\C$ for
$\N$ and $\C$, respectively.

 Green${}^{12}$ has given a class of representations of the above commutation
relations of parafermi creation and annihilation operators. In the so-called 
{\it Green representation\/} for the cases of
order $p$, the parafermi operators $b_k$ are expressed by the form
\[b_k=\sum_{\alpha =1}^pb_k^{(\alpha)},\]
where the {\it Green-component\/} operators $b_k^{(\alpha)}$ satisfy the
commutation relations
\[\{b_k^{(\alpha)},b_l^{(\alpha)\dag}\}=\delta_{k\,l},\,\,\,\,\{b_k^{(\alpha)},
b_l^{(\alpha)}\}=0,\]
\[[b_k^{(\alpha)},b_l^{(\beta)\dag}]=[b_k^{(\alpha)},b_l^{(\beta)}]=0,
\,\,\,\,(\alpha\neq\beta)\]
where $\{A,B\}=AB+BA$, and the uniqueness of vacuum $|0\rangle$ such that
\[b_k^{(\alpha)}|0\rangle=0\,\,\,\,\,\,{\rm for\,\,\, all}\,\,\, k,\alpha.\] 

Green representation is essentially equivalent to the tensor product
representation of Clifford algebra representation of $so(2\nu)$.
In fact, we easily verify that $e_1,...,e_{2\nu}$ defined by $e_{2k-1}=
i(b_k^\dagger+b_k),e_{2k}=b_k^\dagger-b_k$ form the generator of a Clifford
algebra (i.e., $e_i^2=-1$ and $e_ie_j=-e_je_i\,\,(i\neq j)$ hold). 
Thus, we can construct a $2^{p\nu}$-dimensional representation of Green 
components by using a spin representation of the Clifford algebra as follows.
Let $V_k^{(\alpha)}\simeq\C^2\,(k=1,...,\nu,\,\alpha=1,...,p)$. Pauli 
matrices $\sigma_{1,k}^{(\alpha)},\,\sigma_{2,k}^{(\alpha)}$ and $\sigma_{3,k}^
{(\alpha)}$ that act on  $V_k^{(\alpha)}$ are represented as
\[\sigma_{1,k}^{(\alpha)}=\pmatrix{0&1\cr1&0},\,\,\,\,
\sigma_{2,k}^{(\alpha)}=\pmatrix{0&-i\cr i&0},\,\,\,\,
\sigma_{3,k}^{(\alpha)}=\pmatrix{1&0\cr 0&-1}.\]
Define $V^{(\alpha)}$ by 
\[V^{(\alpha)}=V^{(\alpha)}_1\otimes\cdots\otimes V_\nu^{(\alpha)},\]
 and define $\hat{\sigma}_{c,k}^{(\alpha)}$ and $\gamma_{i}^{(\alpha)}
(k=1,...,\nu,i=1,...,2\nu)$ that acts on $V^{(\alpha)}$ by
\[\hat{\sigma}_{c,k}^{(\alpha)}=\overbrace{1\otimes\cdots\otimes1}^{k-1}
\otimes{\sigma_{c,k}^{(\alpha)}}\otimes 1\otimes
\cdots\otimes 1,\,\,\,\,\,c=1,2,3,\]
\[\gamma_{2k-1}^{(\alpha)}=\hat{\sigma}_{2,k}^{(\alpha)}\hat{\sigma}_{3,k+1}
^{(\alpha)}\cdots\hat{\sigma}_{3,\nu}^{(\alpha)},\]
\[\gamma_{2k}^{(\alpha)}=-\hat{\sigma}_{1,k}^{(\alpha)}\hat{\sigma}_{3,k+1}
^{(\alpha)}\cdots\hat{\sigma}_{3,\nu}^{(\alpha)}.\]
Operators $b_k^{(\alpha)}\,\,(k=1,...,\nu)$ defined by
\[b_k^{(\alpha)}=\frac 1{2}(\gamma_{2k-1}^{(\alpha)}-i\gamma_{2k}^{(\alpha)})\]
satisfy the relations
\[\{b_k^{(\alpha)},b_l^{(\alpha)}\dag\}=\delta_{k\,l},\,\,\,\,
\{b_k^{(\alpha)},b_l^{(\alpha)}\}=0,\]
for all $k,\,l=1,...,\nu$.

Define $V$ and $\tilde{b}_k^{(\alpha)}\,(k=1,...,\nu)$ acting on $V$ by
\[V=V^{(1)}\otimes \cdots\otimes V^{(p)},\]
\[\tilde{b}_k^{(\alpha)}=\overbrace{1\otimes\cdots\otimes 1}^{\alpha-1}
\otimes b_k^{(\alpha)}\otimes 1\otimes\cdots\otimes 1.\]
We see that for all $k,l=1,...,\nu$,
\[\{\tilde{b}_k^{(\alpha)},\tilde{b}_l^{(\alpha)\dagger}\}=\delta_{k\,l},\,\,
\,\,\{\tilde{b}_k^{(\alpha)},\tilde{b}_l^{(\alpha)}\}=0,\]
\[[\tilde{b}_k^{(\alpha)},\tilde{b}_l^{(\beta)\dagger}]=
[\tilde{b}_k^{(\alpha)},\tilde{b}_l^{(\beta)}]=0,\,\,\,\,\,(\alpha\neq\beta).
\]
Let $|0\rangle _k^{(\alpha)}\in V_k^{(\alpha)}$ denote the normalized vector
such that $b_k^{(\alpha)}|0\rangle_k^{(\alpha)} =0$. Define $|0\rangle 
^{(\alpha)}$ and $|0\rangle$ by
\[|0\rangle ^{(\alpha)}=|0\rangle _1^{(\alpha)}\otimes\cdots\otimes
|0\rangle _{\nu}^{(\alpha)},\]
\[|0\rangle =|0\rangle ^{(\alpha)}\otimes\cdots\otimes|0\rangle ^{(p)}.\]
Now, we find that $\tilde{b}_1^{(\alpha)},...,\tilde{b}_\nu ^{(\alpha)}$
are $2^{p\nu}$-dimensional representations of Green components and 
$|0\rangle$ is the vacuum. Thus, $b_k=\sum_{\alpha =1}^p\tilde{b}_k^{(\alpha)},
\,\,(k=1,...,\nu)$ are $2^{p\nu}$-dimensional representations of annihilation
operators of $\nu$ parafermi oscillators of order $p$. Let us call the above 
representation of the algebra of parafermi oscillators {\it spin
representation}.

Define $\sigma_{\pm,k}^{(\alpha)}$ by 
\[\sigma_{\pm,k}^{(\alpha)}=
(\sigma_{1,k}^{(\alpha)}\pm i\sigma_{2,k}^{(\alpha)})/2.\]
and $|1\rangle _k^{(\alpha)}\in V_k^{(\alpha)}$ by 
\[|1\rangle _k^{(\alpha)}=\sigma_{+,k}^{(\alpha)}|0\rangle _k^{(\alpha)}.\]
\[\{(|e_1^{(1)}\rangle_1^{(1)}\cdots|e_\nu^{(1)}\rangle_\nu^{(1)})\cdots
(|e_1^{(p)}\rangle_1^{(p)}\cdots|e_\nu^{(p)}\rangle_\nu^{(p)}):
e_k^{(\alpha)}=0,\,1\}\]
($\otimes$'s are omitted) is a complete orthonormal system of $V$. We write
the vectors simply as $|\{e_k^{(\alpha)}\}\rangle$.

Number operator $N$ on $V$  and the related operators $N_k, N^{(\alpha)}$
are defined as follows:
\[N_k^{(\alpha)}=\frac 1{2}(1+\sigma_{3,k}^{(\alpha)})=\pmatrix{1&0\cr 0&0},\]
\[\hat{N}_k^{(\alpha)}=\overbrace{1\otimes\cdots\otimes1}^{k-1}\otimes
N_k^{(\alpha)}\otimes\overbrace{1\otimes\cdots\otimes 1}^{\nu -k},\]
\[\tilde{N}_k^{(\alpha)}=\overbrace{1\otimes\cdots\otimes1}^{\alpha -1}\otimes
\hat N_k^{(\alpha)}\otimes \overbrace{1\otimes\cdots\otimes 1}^{p-\alpha},\]
\[N_k=\sum_{\alpha=1}^p\tilde N_k^{(\alpha)},\,\,\,N^{(\alpha)}=\sum_{k=1}
^\nu\tilde{N}_k^{(\alpha)},\,\,\,N=\sum_{\alpha =1}^pN^{(\alpha)},\]
We see that 
\[N|\{e_k^{(\alpha)}\}\rangle=n|\{e_k^{(\alpha)}\}\rangle\]
where $n$ is the number of $e_k^{(\alpha)}$'s that is equal to 1. It is easily
shown that
\[\tilde b_k^{(\alpha)\dagger}\tilde b_k^{(\alpha)}=\tilde N_k^{(\alpha)},
\,\,\,\,
\tilde b_k^{(\alpha)}\tilde b_k^{(\alpha)\dagger}=1-\tilde N_k^{(\alpha)},
\,\,\,\,
N_k=\frac1{2}([b_k^\dagger,b_k]+p),\]
\[[N_k,N_l]=0,\,\,\,\,\,N_kb_k=b_k(N_k-1),\,\,\,N_kb_k^\dagger=b_k^\dagger
(N_k+1),\,\,\,{\rm etc.}\]

{\it Lemma 5.1:} Suppose that hyperfinite parafermi annihilation 
operators $b_1,...,b_\nu$ are represented by spin representation, and that 
their order $p$ is an infinite hypernatural number( $b_1,...,b_\nu$ are 
$2^{p\nu}\times 2^{p\nu}$ internal matrices acting on $\us\C^{2^{p\nu}}$). 
If $|\xi\rangle\in\us\C^{2^{p\nu}}$ satisfies $\langle\xi |\xi\rangle,\,
\langle\xi |N^2|\xi\rangle <\infty$, and $k\neq l\,(k,l=1,2,...,\nu)$, then
\begin{list}{}{}
\item (i) $[\beta_{k},\beta_{l}]|\xi\rangle\approx [\beta_{k},
\beta_{l}\dag]|\xi\rangle\approx 0.$
\item (ii) $[\beta_{k},\beta_{k}\dag]|\xi\rangle\approx|\xi\rangle.$
\item (iii) $\beta_k\beta_k^{\dagger n}|\xi\rangle\approx
(\beta_k^{\dagger n}\beta_k
+n\beta_k^{\dagger n-1})|\xi\rangle$.
\end{list}
where $\beta_k=p^{-1/2}b_k$ ({\it normalization\/} of $b_k$) 
and $n<\infty$.

{\it Proof:}
(i) Notice that 
\[[\beta_{k},\beta_{l}]=\frac2{p}\sum_{\alpha=1}^p\tilde{b}_k^{(\alpha)}
\tilde{b}_l^{(\alpha)},\,\,\,\,\,
[\beta_{k},\beta_{l}\dag]=\frac2{p}\sum_{\alpha=1}^p\tilde{b}_k^{(\alpha)}
\tilde{b}_l^{(\alpha)\dagger}.\]
Direct calculations show that
\[b_k^{(\alpha)}b_l^{(\alpha)}=\cases{-\hat\sigma_{3,k+1}^{(\alpha)}\cdots
\hat\sigma_{3,l}^{(\alpha)}\hat\sigma_{-,k}^{(\alpha)}
\hat\sigma_{-,l}^{(\alpha)} & ($l>k$)\cr
-\hat\sigma_{3,l+1}^{(\alpha)}\cdots\hat\sigma_{3,k}^{(\alpha)}
\hat\sigma_{-,k}^{(\alpha)}\hat\sigma_{-,l}^{(\alpha)} & ($k>l$)\cr}\]
\[b_k^{(\alpha)}b_l^{(\alpha)\dag}=\cases{\hat\sigma_{3,k+1}^{(\alpha)}\cdots
\hat\sigma_{3,l}^{(\alpha)}\hat\sigma_{+,k}^{(\alpha)}
\hat\sigma_{-,l}^{(\alpha)} &($l>k$)\cr
\hat\sigma_{3,l+1}^{(\alpha)}\cdots\hat\sigma_{3,k}^{(\alpha)}
\hat\sigma_{+,k}^{(\alpha)}\hat\sigma_{-,l}^{(\alpha)} & ($k>l$)\cr}\]
Hence, using $N_k^{(\alpha)}+N_l^{(\alpha)}-2N_k^{(\alpha)}N_l^{(\alpha)}
=(N_k^{(\alpha)}-N_l^{(\alpha)})^2\geq 0$,
\[\|\tilde{b}_k^{(\alpha)}\tilde{b}_l^{(\alpha)}|\xi\rangle\|^2
=\langle\xi|\tilde{\sigma}_{+,k}^{(\alpha)}\tilde{\sigma}_{+,l}^{(\alpha)}
\tilde{\sigma}_{-,l}^{(\alpha)}\tilde{\sigma}_{-,k}^{(\alpha)}|\xi\rangle\]
\[\hspace{3em}=\langle\xi|\tilde N_k^{(\alpha)}\tilde N_l^{(\alpha)}|
\xi\rangle\]
\[\hspace{5em}\leq\frac 1{2}\langle\xi|\tilde N_k^{(\alpha)}+\tilde N_l
^{(\alpha)}|\xi\rangle\]
\[\hspace{2em}\leq\frac 1{2}\langle\xi |\tilde N^{(\alpha)}|\xi\rangle.\]
Thus,
\[\|[\beta_{k},\beta_{l}]|\xi\rangle \|=\|\frac2{p}\sum_{\alpha=1}^p
\tilde{b}_k^{(\alpha)}\tilde{b}_l^{(\alpha)}|\xi\rangle \|\]
\[\hspace{6em}\leq\frac2{p}\sum_{\alpha=1}^p\|\tilde{b}_k^{(\alpha)}
\tilde{b}_l^{(\alpha)}|\xi\rangle \|\]
\[\hspace{9em}\leq\frac2{p}\sum_{\alpha=1}^p\sqrt{\frac1{2}\langle\xi|
\tilde N^{(\alpha)}|\xi\rangle}\approx 0.\]
The last equation is seen from $\sum_{\alpha =1}^p\langle\xi|
N^{(\alpha)}|\xi\rangle =\langle\xi|N|\xi\rangle <\infty$; under this 
condition we find that the number of $\alpha$'s such that $\langle\xi|
N^{(\alpha)}|\xi\rangle\not\approx 0$ is finite. 

In a similar way, it is shown that
\[\|[\beta_{k},\beta_{l}\dag]|\xi\rangle \|\leq\frac2{p}\sum_{\alpha=1}^p
\|\tilde{b}_k^{(\alpha)}\tilde{b}_l^{(\alpha)}\dag|\xi\rangle \|\] 
\[\hspace{40mm}\leq\frac2{p}\sum_{\alpha=1}^p\sqrt{\langle\xi |\tilde N_l
^{(\alpha)}(1-\tilde N_k^{(\alpha)})|\xi\rangle}\]
\[\hspace{30mm}\leq\frac2{p}\sum_{\alpha=1}^p\sqrt{\langle\xi |\tilde N_l
^{(\alpha)}|\xi\rangle}\approx 0.\]

(ii) A direct calculation shows that
\[[\beta_{k},\beta_{k}\dag]=-\frac1{p}\sum_{\alpha=1}^p\tilde\sigma_{3,k}
^{(\alpha)},\]
\[([\beta_{k},\beta_{k}\dag]-1)^2=\frac4{p^2}\left(\sum_{\alpha=1}^pN_k
^{(\alpha)}\right)^2.\]
Hence, 
\[\|[\beta_{k},\beta_{k}\dag]|\xi\rangle -|\xi\rangle\|^2
=\frac4{p^2}\langle\xi|\left(\sum_{\alpha=1}^pN_k^{(\alpha)}\right)|\xi\rangle.
\]
\[\hspace{40mm}\leq\frac4{p^2}\langle\xi|N^2|\xi\rangle\approx 0.\]

(iii) is shown by the induction for $n$, using (ii). Q.E.D.

Suppose the number of parafermi oscillators $\nu$ and their order $p$
are infinite hypernatural numbers. When $n_i$ is nonnegative integer for any 
$i=1,2,...<\infty$, and the number of $n_i$'s such that $n_i\neq 0$ is 
finite, we will define $|n_1,n_2,...\rangle$ by
\[|n_1,n_2,...\rangle=\frac{b_{1}^{\dagger n_1}b_{2}^{\dagger n_2}
\cdots|0\rangle}{\|b_{1}^{\dagger n_1}b_{2}^{\dagger n_2}
\cdots|0\rangle\|}.\]
$b_{1}^{\dagger n_1}b_{2}^{\dagger n_2}\cdots$ is the product of a 
finite number of operators, so it is well-defined. $N_k|n_1,n_2,...\rangle
=n_k|n_1,n_2,...\rangle$ is easily shown, and hence, since $N_k$ is hermitian,
the set of the vectors of the form $|n_1,n_2,...\rangle$ is an orthonormal
system.

{\it Lemma 5.2:}
\begin{list}{}{}
\item (i) $\beta_k^\dagger\beta_k|n_1,n_2,...\rangle\approx n_k
|n_1,n_2,...\rangle$,
\item (ii) $\beta_k\beta_k^\dagger|n_1,n_2,...\rangle\approx (n_k+1)
|n_1,n_2,...\rangle$,
\item (iii) $\|\beta_1^{\dagger n_1}\beta_2^{\dagger n_2}\cdots |0\rangle \|
\approx \sqrt{n_1!n_2!\cdots}$,
\item (iv) $\beta_k^\dagger|n_1,n_2,...\rangle\approx\sqrt{n_k+1}|n_1,n_2,...,
n_k+1,...\rangle$,
\item (v) $\beta_k|n_1,n_2,...\rangle\approx \sqrt{n_k}|n_1,n_2,...,
n_k-1,...\rangle$.
\end{list}

{\it Proof:}
(i) and (ii) are shown by Lemma 5.1(i)-(iii). (iii) is shown by
the induction for $n_1,n_2,...$, using (ii) and Lemma 5.1(i). (iv) and (v)
are shown by (iii) and Lemma 5.1(i)(iii). Q.E.D.

Define a set $D\subset \us\C^{2^{p\nu}}$ by
\[D=\Bigl\{ \frac{\beta_{k_1}^\dagger\cdots\beta_{k_n}^\dagger|0\rangle}
{\|\beta_{k_1}^\dagger\cdots\beta_{k_n}^\dagger|0\rangle\|}\Big|\,n,k_1,...,
k_n\in\N\Bigr\}\cup\{|0\rangle\}.\] 
Clearly, every vector in $D$ is a normalized eigenvector of the number 
operator $N$
with a finite eigenvalue. Let $S$ denote the external subspace of 
$\us\C^{2^{p\nu}}$ spanned by $D$, i.e.,
\[S=\{\sum_{i=1}^nc_i|\xi\rangle\,:\,\,c_i\in\us\C, |c_i|<\infty, n\in\N,
|\xi\rangle\in D\}.\]
The following theorem follows from Lemma 5.1 and 5.2.

{\bf Theorem 5.3}: $(\beta_k,S)\,\,(k\in\N)$ is a representation of CCR
of countably-infinite degree of freedom, i.e., $S$ is invariant in respect to 
$\beta_k$ and $\beta_k^\dagger$ for every $k\in\N$, and 
\[[\beta_k,\beta_l]|\xi\rangle\approx 0\]
\[[\beta_k,\beta_l^\dagger]|\xi\rangle\approx \delta_{kl}|\xi\rangle\]
for any $|\xi\rangle\in S$. Moreover, the uniqueness of vacuum is 
satisfied in the following sense: if $|\xi\rangle\in S$, $\langle\xi|
\xi\rangle=1$ and 
$\beta_k|\xi\rangle\approx 0$ for all $k\in\N$, then $|\langle\xi|0
\rangle|\approx 1$.

\section*{APPENDIX: NONSTANDARD ANALYSIS}
 
This section briefly outlines the basic notions of nonstandard analysis 
along Hurd and Loeb${}^6$, and reviews some development in the field of 
hyperfinite-dimensional linear space theory.
Let $X$ be a set and ${\cal P}(X)$ the power set of $X$, that is, the set 
of all subsets of $X$. The {\it superstructure over} $X$, denoted by $V(X)$,
 is 
defined by the following recursion:
\[V_0(X)=X, V_{n+1}(X)=V_n(X)\cup{\cal P}(V_n(X)),\]
\[V(X)=\bigcup _{n\in {\bf N}}V_n(X).\]
Let us regard any element of 
$X$ as a nonset here; hence $x\in V(X)$ is a set iff $x\in V(X)\setminus X$. 
Let 
${\bf C}$ be the set of complex numbers. $V({\bf C})$ contains all the 
structures that we usually use in quantum physics, for instance, separable 
Hilbert space ${\cal H}$.

$V(X)$ is called a {\it nonstandard extension of\/} $V({\bf C})$
 if there exists a map $\star :V({\bf C})\longrightarrow V(X)$ satisfying 
the following conditions:
\begin{list}{}{}
\item (1)  $\star $ is an injective mapping from $V({\bf C})$ to $V(X)$,
\item (2)  $^\star {\bf C}=X$,
\item (3) (Transfer Principle) Let $\phi$ be a sentence in terms of 
$V({\bf C})$, and $^\star\phi $ the sentence ``transfered'' from $\phi$ 
by mapping $\star.$ $\phi$ is true iff $^\star\phi$ is true.
\end{list}

\begin{sloppypar}
Transfer Principle needs more explanation. Although the exact description of it
needs more definitions of some notions in mathematical logic, 
one can find the intuitive meaning of it without them, in an example. A 
sentence in terms of $V({\bf C})$
 is constructed from the symbols for logical connectives $\neg, \wedge, \vee, 
\Rightarrow, \Leftrightarrow$, quantifiers $\forall, \exists$, individual 
variables $x,y,z,...$, two predicates $=,\in$, parentheses $(, )$, and 
elements of $V({\bf C})$. We will consider an example. 
Let {\bf R} denote the set of real 
numbers. Define $G_<\in V({\bf C})$ by 
$G_<=\{(x,y)|\,x,y\in {\bf R}, x<y\}$, where $(x,y)$ 
is identified as $\{\{x\},y\}$. 
\end{sloppypar}
\[(\forall x)(\forall y)(x\in{\bf R}\wedge y\in{\bf R}\wedge (x,y)\in G_<
\Rightarrow(\exists z)(z\in{\bf R}\]
\[\wedge (x,z)\in G_<\wedge (z,y)\in G_<))\]
 is a sentence in terms of $V({\bf C})$ because ${\bf R},G_<\in V({\bf C})$.
 Let $\phi$ denote this sentence. $\phi$ means that ${\bf R}$ is dense, 
and hence $\phi $ is true. The ``transfered'' sentence $^\star\phi$ is as 
follows:
\[(\forall x)(\forall y)(x\in\!^\star{\bf R}\,\wedge \,y\in\!^\star{\bf R}
\,\wedge\,(x,y)\in\! ^\star G_<\Rightarrow(\exists z)(z\in\!^\star{\bf R}\]
\[\wedge\, (x,z)\in\! ^\star G_<\,\wedge \,(z,y)\in\! ^\star G_<))\]
By Transfer Principle, $^\star\phi$ is true  ($^\star{\bf R}$ is called 
$\star$-dense).

$u\in V(X)$ is called {\it standard\/} if there is $x\in V({\bf C})$ such that
 $u=$$^\star x$, and called {\it internal\/} if there is $x\in V({\bf C})$ 
such 
that $u\in$$^\star x$. $^\star V({\bf C})$ is the set of all internal sets. 
Let $A$ and $B$ be internal sets.
Function $f:A\longrightarrow B$ is called {\it internal\/} if the graph of $f$
(i.e., $\{(x,f(x))|x\in A\}$) is internal. $V(X)$ is called a {\it countably 
saturated extension of\/} 
$V({\bf C})$ if it satisfies the following condition:

(Saturation Principle) If countable sequence of internal sets $A_j\in V(X)
\setminus X$ satisfies
\[\bigcap_{j=1}^kA_j\neq\phi\,(k=1,2,...)\]
then
\[\bigcap_{j=1}^{\infty}A_j\neq\phi.\]

Let $A\in V({\bf C})\setminus {\bf C}$. From the Saturation Principle, we can 
show that if $A$ is an infinite set, $^sA$ defined by 
\[^sA=\{^\star a|\,a\in A\}\]
is a proper subset of $^\star A$. Hence $^\star A$ is an extended structure
 of $A$. 
$^\star A$ is called the {\it nonstandard extension of\/} A. 
By renaming the elements of $X$,
 we can assume without loss of generality that ${\bf C}$ is a subset of $^\star
 {\bf C}$ and $^\star x=x$ for each $x\in {\bf C}$.

Given any subset $U\subseteq V(\C)$, define $^\star U$ by
\[^\star U=\bigcup_{n\in {\bf N}}\!^\star(U\cap V_n(\C)).\]
Let $F({\bf C})\subseteq V({\bf C})$ be the set of finite sets, i.e.,
\begin{center}
$F({\bf C})=\{A\in V({\bf C})\setminus{\bf C}|\,A$ is a finite set\}.
\end{center}
An element of $^\star F({\bf C})$ is called a {\it hyperfinite set\/}. 
It is shown that if $A$ is a
hyperfinite set, then there
is an initial segment $J=\{n\in\!^\star{\bf N}|\,n\le j\}$ for some $j\in\!
^\star {\bf N}$ and a one-to-one, onto internal mapping $f:J\longrightarrow A$.
Thus, we will often write a hyperfinite set $A$ as $A=\{a_1,a_2,...,a_j\}
 $, where $a_k=f(k),k\in J$.

Any element of $^\star {\bf C}\,(^\star {\bf R})$ is called a {\it 
hypercomplex (hyperreal)\/} number. We assume that $^\star 
{\bf R}\subseteq\!^\star {\bf C}$.
 $^\star{\bf C}$ is a proper extension field of ${\bf C}$, 
 and $^\star {\bf R}$ is an ordered extension of ${\bf R}$.
An element of $^\star {\bf N}\,(^\star {\bf Z})$ is called a {\it 
hypernatural number (hyperinteger)\/}. 
A hypercomplex number $x$ is called {\it infinite\/} if $|x|>n$ for any $n\in 
{\bf N}$, {\it finite\/}, $|x|<\infty$, if there is some $n\in {\bf N}$ such
 that $|x|<n$, and {\it infinitesimal\/} if $|x|<\frac{1}{n}$ for any $n\in
{\bf N}$.

For any $x,y\in \!^\star {\bf C}$, we will write $x\approx y$ if $|x-y|$ is
infinitesimal. For any finite hypercomplex number $x$, there is a unique 
complex number $z$ such that $^\star z\approx x$; this $z$ is called the 
{\it standard part of\/} of $x$ and denoted by $^\circ x$.

\def\fin{\mathop{\rm fin}\nolimits}
Let $\cal A$ be an internal normed linear space with norm $\parallel \cdot
\parallel$. The {\it principal galaxy\/} $\fin({\cal A})$ and the {\it 
principal monad\/} $\mu(0)$ are defined by
\[\fin({\cal A})=\{x\in{\cal A}|\,\parallel x\parallel<\infty\}\]
\[\mu(0)=\{x\in{\cal A}|\,\parallel x\parallel \approx 0\}\]
 Both of them are linear spaces over $\C$. The {\it nonstandard hull of\/}
${\cal A}$ is the quotient linear space $\hat{{\cal A}}=\fin({\cal A})/\mu(0)$
equipped with the norm given by
\[\parallel ^\circ x\parallel=\!^\circ \parallel x\parallel\]
for all $x\in \fin({\cal A})$, where $^\circ x=x+\mu(0)$. It is shown
by the Saturation Principle that $\hat{\cal A}$ is a Banach space${}^6$.

Let $\nu$ be an infinite hypernatural number, i.e., $\nu\in\! ^\star \N
\setminus
\N$. Define $\us\C^\nu$ by
\[\us\C^\nu=\{f|\,f:\{1,...,\nu\}\rightarrow\us\C;\, {\rm internal}\}.\]
$^\star\C^\nu$ is a $\nu$-dimensional internal inner product space
with the natural inner product and the internal norm $\parallel\cdot\parallel
$ derived by the inner product.
Define $\M=M(\us\C,\nu)$, the set of $\nu\times\nu$ internal matrices 
over $\us\C$, by
\[\M=\{f|\,f:\{1,...,\nu\}^2\rightarrow \us\C;\,{\rm internal}\}.\]
Naturally $\M$ acts
on $^\star\C^\nu$ as the internal linear operators. Let $p_{\infty}$
be the operator norm on $\M$, i.e., $p_{\infty}(A)=\sup\{\parallel A\xi
\parallel |\,\parallel\xi\parallel\le 1, \xi\in\!^\star\C^\nu\}$. Denote
by $A^*$ the adjoint of $A\in\M$. Let $\tau$ be the internal normalized
trace on $\M$, i.e., 
\[\tau (A)=\frac{1}{\nu}\sum_{i=1}^{\nu}A_{ii}\]
for $A=(A_{ij})\in\M$. Then $\tau$ defines an internal inner product
$(\cdot|\cdot)$ on $\M$ by $(A|B)=\tau(A^*B)$, for $A,B\in\M$. Its derived
norm called the normalized Hilbert-Schmidt norm is denoted by $p_2$, i.e.,
$p_2(A)=\tau(A^*A)^{1/2}$ for $A\in \M$. 
Denote by $(\M,p_\infty)$ and $(\M,p_2)$ the normed linear spaces equipped
with these respective norms. The principal galaxies $\fin_{\infty}(\M)$ of
$(\M,p_{\infty})$ and $\fin_2(\M)$ of $(\M,p_2)$ are defined as follows:
\[\fin_{\infty}(\M)=\{A\in\M|\,p_{\infty}(A)<\infty\}\]
\[\fin_{2}(\M)=\{A\in\M|\,p_{2}(A)<\infty\}\]
The principal monads $\mu_{\infty}(0)$ of $(\M,p_{\infty})$ and $\mu_2(0)
$ of $(\M,p_2)$ are defined as follows:
\[\mu_{\infty}(0)=\{A\in\M|\,p_{\infty}(A)\approx  0\}\]
\[\mu_{2}(0)=\{A\in\M|\,p_{2}(A)\approx  0\}.\]
The nonstandard hull $\hat\M_2=\fin_2(\M)/\mu_2(0)$ turns out to be a Hilbert
space with inner product $<\cdot|\cdot>$ and norm $\parallel\cdot\parallel
_2$ defined by 
\[<A+\mu_2(0)|B+\mu_2(0)>=\!^{\circ}(A|B)\]
and
\[\parallel A+\mu_2(0)\parallel_2=\!p_2(A).\]
for $A,B\in\fin_2(\M)$.

The nonstandard hull $\hat{\M}_{\infty}=\fin_{\infty}(\M)/\mu_{\infty}(0)$
of $(\M,p_{\infty})$ turns out to be a $C^*$-algebra equipped with norm
$\hat p_{\infty}$ defined by 
\[\hat p_{\infty}(A+\mu_{\infty}(0))=\!^\circ p_{\infty}(A)\]
 for $A\in\fin_{\infty}(\M)$.

Hinokuma and Ozawa${}^{14}$ showed that another quotient space $\hat\M$ 
defined by
\[\hat\M=\fin_{\infty}(\M)/(\mu_2(0)\cap\fin_{\infty}(\M)).\]
is a von Neumann algebra of type I\rm{I}$_1$ factor.

\section*{ACKNOWLEDGEMENTS}
The author expresses his appreciation to Professor M. Ozawa for his genaral
supports. 
\vspace{10mm}\\  
${}^1$I.Ojima and M.Ozawa, ``Unitary Representations of the Hyperfinite
Heisenberg Group and the Logical Extension Methods in Physics'', Open
Systems and Information Dynamics. 2(1),107-128(1993).\\
${}^2$S.Gudder, ``Toward a Rigorous Quantum Field Theory,''
Found.Phys. 24(9), 1205-1225(1994).\\
${}^3$M.Ozawa, ``Phase operator problem and macroscopic extension of 
quantum mechanics,'' preprint (Nagoya University, 1992).\\
${}^4$T.Nakamura, ``A Nonstandard Representation of Feynman's
Path Integrals,'' J.Math.Phys. 32(2), 457-463(1991).\\
${}^5$A.Robinson, {\it Non Standard Analysis}(North-Holland,
Amsterdam,1966).\\
${}^6$A.E.Hurd, and P.A.Loeb, {\it An Introduction to Nonstandard
Real Analysis}(Academic Press,Orlando,1985).\\
${}^7$S.Gudder and V.Naroditsky, ``Finite-Dimensional Quantum 
Mechanics,'' Int.J.Theor.Phys. 20(8),619-643(1981).\\
${}^8$A.O.Barut and A.J.Bracken, ``Compact Quantum systems: Internal
Geometry of Relativistic Systems,'' J.Math.Phys. 26(10),2515-2519(1985).\\
${}^9$R.J.B.Fawcett and A.J.Bracken, ``Simple Orthogonal and Unitary
Compact Quantum Systems and the In\"on\"u-Wigner Contraction,'' J.Math.
Phys.29(7),1521-1528(1988).\\
${}^{10}$D.T.Pegg and S.M.Barnett, ``Phase properties of the quantized
single-mode electromagnetic field,'' Phys.Rev.A 39(4),1665-1675(1989).\\
${}^{11}$F.T.Arecchi, et al., ``Atomic Coherent States in Quantum 
Optics,'' Phys.Rev.A 6(6),2211-2237(1972).\\
${}^{12}$H.S.Green, ``A generalized method of field quantization,'' 
Phys.Rev. 90(2),270(1953).\\
${}^{13}$Y.Ohnuki and S.Kamefuchi, {\it Quantum Field Theory and 
Parastatistics\/} (University of Tokyo Press,Tokyo,1982).\\
${}^{14}$T.Hinokuma and M.Ozawa, ``Conversion from Nonstandard Matrix 
Algebras to Standard Factors of Type II$_1$,'' Illinois.J.Math.
37(1),1--13(1993).\\
${}^{15}$L.C.Moore,Jr., ``Hyperfinite Extensions of Bounded Operators on a
Separable Hilbert Space,'' Trans.Am.Math.Soc. 218,285-295(1976).

\end{document}